\documentclass[twocolumn,nofootinbib,jcp,preprintnumbers,amsmath,amssymb]{revtex4-1}
\usepackage{natbib,hyperref}
\usepackage{amsmath}
\usepackage{graphicx}
\usepackage{dcolumn}
\usepackage{bm}
\usepackage{epsfig,color,xspace,multirow,xr,bbold}
\usepackage[all]{xy}
\usepackage{setspace}
\usepackage{url}
\usepackage[colorinlistoftodos]{todonotes}
\usepackage{threeparttable}

\usepackage{natbib,hyperref}
\usepackage{float}
\usepackage{placeins}
\usepackage[utf8]{inputenc}

\newcommand{\INBLUE}[1]{\textcolor{black}{#1}}

\usepackage{xcolor, soul}
\sethlcolor{green}
\usepackage[none]{hyphenat}

\begin{document}

\title{
High-Throughput Design of Peierls and Charge Density Wave Phases \\
in Q1D Organometallic Materials
}
\date{\today}     

\author{Prakriti Kayastha$^{1}$}
\author{Raghunathan Ramakrishnan$^{1}$}

\email{ramakrishnan@tifrh.res.in}

\affiliation{$^1$Tata Institute of Fundamental Research, Centre for Interdisciplinary Sciences, Hyderabad 500107, India}


\begin{abstract}
Soft-phonon modes of an undistorted phase 
encode a material's preference for symmetry lowering. 
However, the evidence is sparse for the 
relationship between an unstable phonon wavevector's reciprocal and the number of formula units in the stable distorted phase.
This ``$1/q^*$-criterion'' holds great potential for the first-principles
design of materials, especially in low-dimension. We validate the approach 
on the Q1D organometallic materials space containing 1199 ring-metal units 
and identify candidates that are stable in undistorted (1 unit), Peierls (2 units), 
charge density wave (3-5 units), or long wave ($>5$ units) phases. 
We highlight materials exhibiting gap-opening as well as an uncommon 
gap-closing Peierls transition, and discuss an example
case stabilized as a charge density wave insulator. 
We present the data generated for this study  
through an interactive publicly accessible Big Data analytics platform
\href{https://moldis.tifrh.res.in/data/rmq1d}{(https://moldis.tifrh.res.in/data/rmq1d)}
facilitating limitless and seamless data-mining explorations. 
\end{abstract}

\maketitle
    Materials stabilized in low-dimension offer maximum tunability of properties through external perturbations---an understanding of their dynamical behavior is crucial for rational materials design\cite{cao2019electromagnetic}. 
    The soft-phonon mode in an undistorted phase steers geometric distortions culminating in modulated superstructures. The corresponding change in the electron density gives rise to a charge density wave (CDW)\cite{gruner1988dynamics}.  
    Peierls transition is a unique case of CDW where the maximally unstable phonon wavevector ($q$) in a Q1D material is at the edge of the first Brillouin zone, ${\rm X}$\cite{peierls1996quantum}. 
    Studies on synthesis and characterization of 
    quasi-one-dimensional (Q1D) organic/organometallic materials are 
    abound in anticipation of observing exceptional CDW  properties. 
    CDW materials exhibit a broad range of anisotropic electric, thermal or interesting dynamical behaviour\cite{casian2010violation,ye2015topological,toombs1978quasi,etemad1982polyacetylene,heeger1988solitons,giamarchi2004theoretical,ahn2004mechanism}.  Some widely characterized Q1D 
    conductors exhibiting CDW stabilization are 
    mixed-valence planar transition metal complexes 
    and transition metal chalcogenides\cite{gopalakrishnan1983transition,hoffmann1980electronic,bao2015superconductivity}, while prominent Q1D insulators are of the
    ABX$_3$-type showing anisotropic magnetism\cite{ackerman1974physical,harrison1991magnetic,endoh1974dynamics}.
    Low-dimensional materials such as carbon nanotubes, nanoribbons\cite{piscanec2007optical,connetable2005room,tozzini2010electronic} and conjugated polymers\cite{he2014finite,tolbert1997beyond} show high thermoelectric performance in their Peierls phase. 
    
\begin{figure}[!hbp]
    \centering
    \includegraphics[width=6.0cm]{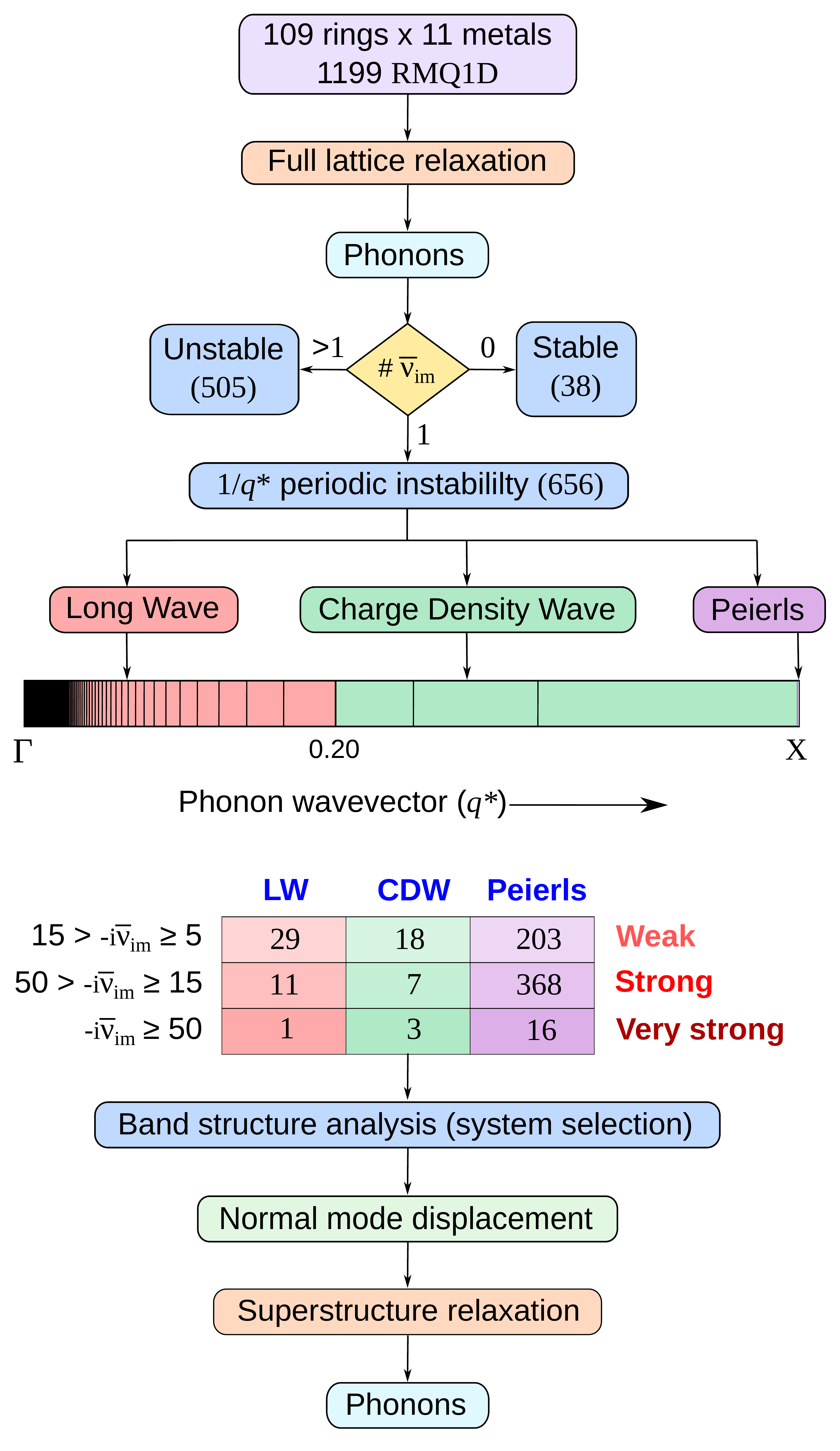}
    \caption{
    Workflow for high-throughput modeling of RMQ1D materials:
    Characterization of dynamic stability is 
    based on the number of imaginary phonon modes 
    (${\overline{\nu}_{\rm im}}$ is the wavenumber) and maximally unstable
    phonon wavevector ($q^*$). Discrete spectrum points 
    to commensurate values of $q^*$.
    Integer-valued $1/q^*$ gives the number of units 
    in Peierls, Charge Density Wave or Long Wave superstructure phases. 
    Numbers reported can be generated on-the-fly by data-mining
    at \href{https://moldis.tifrh.res.in/data/rmq1d}{(https://moldis.tifrh.res.in/data/rmq1d)}; see Supplementary Information (SI) for query screenshots.
    }
    \label{fig:fig1}
\end{figure}

    In this study, we present a strategy for the rational design of 
    dynamically stable phases of Q1D materials using information encoded
    in the phonon band structure of the undistorted, translationally 
    symmetric phase.
Our approach builds on the fact that the number of units preferred 
 in low symmetric phase corresponds to maximal instability at $q^*=1/N$; for integer $N$, we find a commensurately modulated phase.  
 Although generally every $N>1$ indicates a CDW, in this study, we refer to the $N=2$ case as Peierls. 
 While Peierls stabilization is associated with a metal-to-insulator transition,
some studies have reported a reduction in translational symmetry through 
transverse displacements without gap-opening at the zone-edge\cite{batra1990gapless,batra1991peierls}. 
   \INBLUE{The $1/q^*$ criterion generalizes
    Peierls and Jahn-Teller-type instabilities---both effects
    marked by the distortion of a symmetric structure.  
    The former is signified by a gap opening at the band's zone-edge, while the latter
    is accompanied by the lifting of degeneracy in the 
    highest occupied molecular orbital---amounting
    to opening a gap in molecular orbitals. Irrespective of the final outcome, 
    i.e., gap-opening at the level of bands or orbitals, the overall structural instability is characteristic of the dynamic stability of a geometry on the material potential energy surface.}
    Hence, the presence of dynamic instability in a system constitutes a necessary condition for superstructure formation, whereas the presence of a metallic band 
    in the undistorted phase, a sufficient one.
 
    To validate our high-throughput strategy based on the $q^*=1/N$ criterion
    for designing translationally low symmetric phases,
    we select  
    1D chain of alternate ring-metal units (RMQ1D)\cite{bohm1987one}. 
    The materials space is composed of
    11 monovalent metals (Na--Cs, Cu--Au, Al--Tl) bonded to any of the
    109 rings generated
    by substituting C atoms in a cyclopentadienyl anion (Cp)
   with all possible valencies of B, N, and S.
    Cp complexes with Li$^+$ and Na$^+$, 
    form an eclipsed chain, while K$^{+}$, In$^{+}$ and Tl$^{+}$ prefer a
 zigzag form\cite{dinnebier1997solid,downs1993chemistry}. 
 The structures are stable because they minimize repulsion due to
 ring-ring interactions as well as 
 interactions between the ring and the distorted core of the metal\cite{canadell1984theoretical, bytheway1996topological}.
Formation of such structures has escaped inspection in the light of  Peierls instability
because of the lack of gap-opening in these materials that remain as insulators before
and after distortion. 
On the other hand, the longitudinal distortion in 
B$_2$C$_3$H$_5$Ni leads to widening of the Ni bands that is
inextricably coupled with the dimerization of the units\cite{siebert1988polydecker,lavrentiev1993theoretical}. 
    Such Q1D materials are unknown for the $d^{10}$ metals, 
and for combinatorially varying heteroatom-substituted Cp. For 
comprehensive coverage and data-driven
analysis, we present an {\it ab initio} high-throughput workflow (Figure~\ref{fig:fig1}). Such an automated 
strategy has gained popularity for the design of  molecular\cite{hachmann2011harvard,ramakrishnan2014quantum,chakraborty2019chemical,senthil2020troubleshooting}
and materials\cite{curtarolo2013high,jain2013commentary,kirklin2015open} Big Data.

Based on the nature of phonon instability, our high-throughput workflow characterizes stationary points on the material potential energy surface into three classes: 
    (i) stable with positive phonons,
    (ii) those with maximum instability in a (degenerate or non-degenerate) 
    normal mode, and 
    (iii) unstable with imaginary frequencies in several modes. 
    Distribution of the 1199 materials into these classes and
further analysis of the stability was performed through data-mining, see Figure~\ref{fig:fig1}. 
With a threshold of 5 cm$^{-1}$ for the imaginary wavenumbers' norms, we find 38 materials belonging to class-i, stabilizing in the 1U phase with
one ring-metal formula unit. In class-ii, we find 656 materials that provide
a scope to apply the $1/q^*$ criterion for stabilizing in a Peierls, CDW, or long wave phase through normal mode displacements followed by superstructure relaxation.
    While our flow characterized 505 RMQ1Ds as unstable (class-iii), with
multiple imaginary phonon modes, one cannot
rule out the possibility of locating a
local minimum on the potential energy surface for these
stoichiometries when starting with a unit cell arrangement with lateral packing.
\begin{figure}[!htp]
    \centering
    \includegraphics[width=8.5cm]{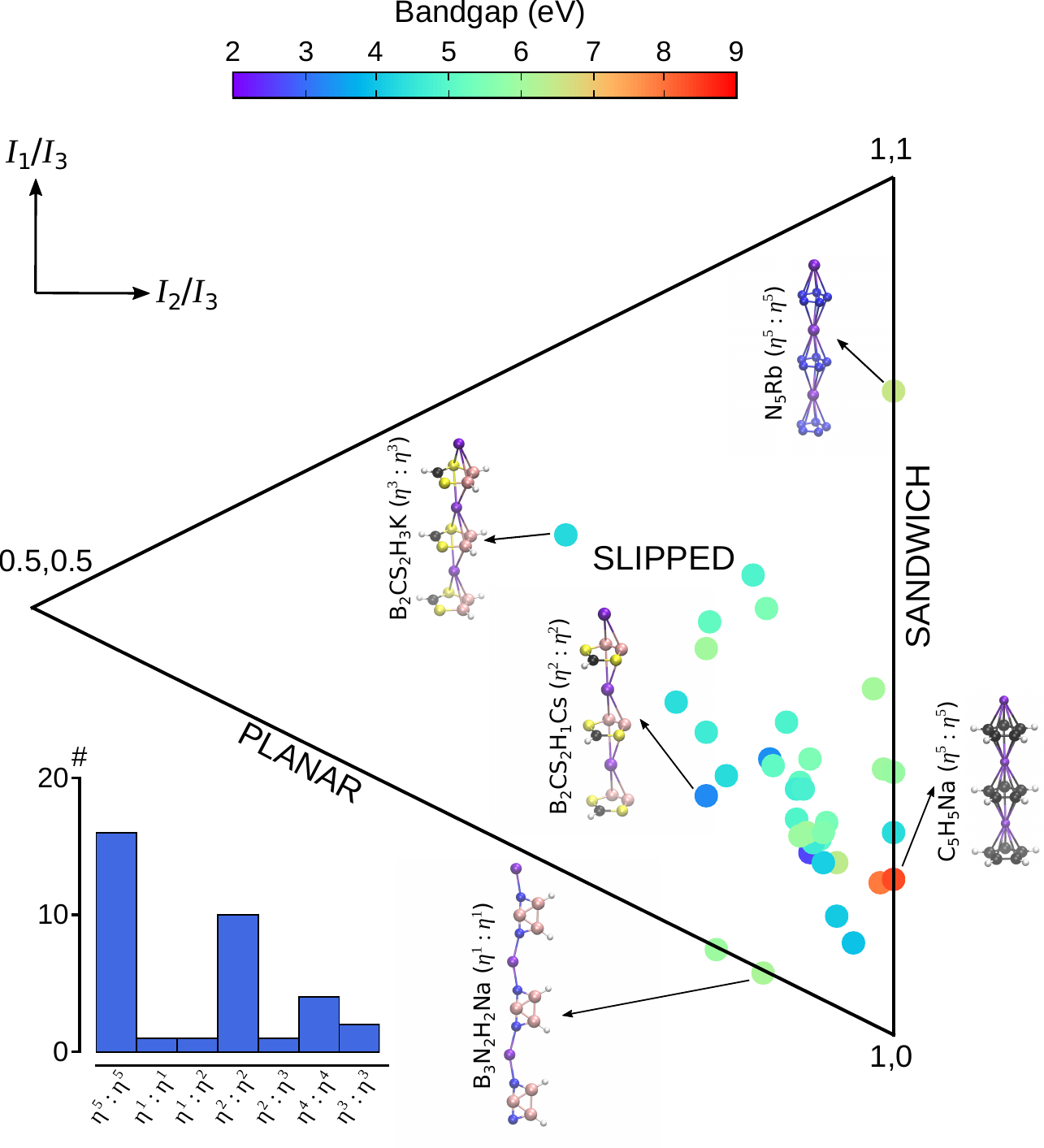}
    \caption{
Shape distribution of RMQ1D materials stable in the 1U phase, with one formula unit. The axes correspond to normalized principal moments of inertia $I_1$/$I_3$ and $I_2$/$I_3$, respectively, 
with $I_1 \le I_2 \le I_3$. 
For 38 dynamically stable materials, PBE0 band gaps are color-coded. The inset shows the distribution of their hapticities ($\eta$). 
    }
    \label{fig:fig2}
\end{figure}

We identify 38 dynamically stable 1U materials with large band gaps, $\varepsilon_{\rm g}\ge2$ eV. Three of these, including CpNa characterized by diffraction\cite{dinnebier1997solid}, exhibit positive phonon wavenumbers 
throughout the Brillouin zone, while for the rest
35, we observed imaginary wavenumbers of small magnitudes ($<5$ cm$^{-1}$). 
The computed value for the longitudinal 
lattice constant of CpNa is 4.74 \AA{}, which is within $<1\%$ error
compared to the experimental value 4.71 \AA{}. 
The most distinguishing feature of the
systems stable in the 1U phase is that they are exclusively
alkali metal-based. 
Figure~\ref{fig:fig2}
provides a bird's-eye view of the structural diversity and band gaps 
of the 38 1U systems.
Materials showing ring-opening or fragmentation 
have been eliminated; these rings exhibit strong $\sigma$-donation to the metal center resulting 
in a loss of bonding electrons needed to stabilize the ring framework. 

\begin{figure*}[!hpt]
    \centering
    \includegraphics[width=15cm]{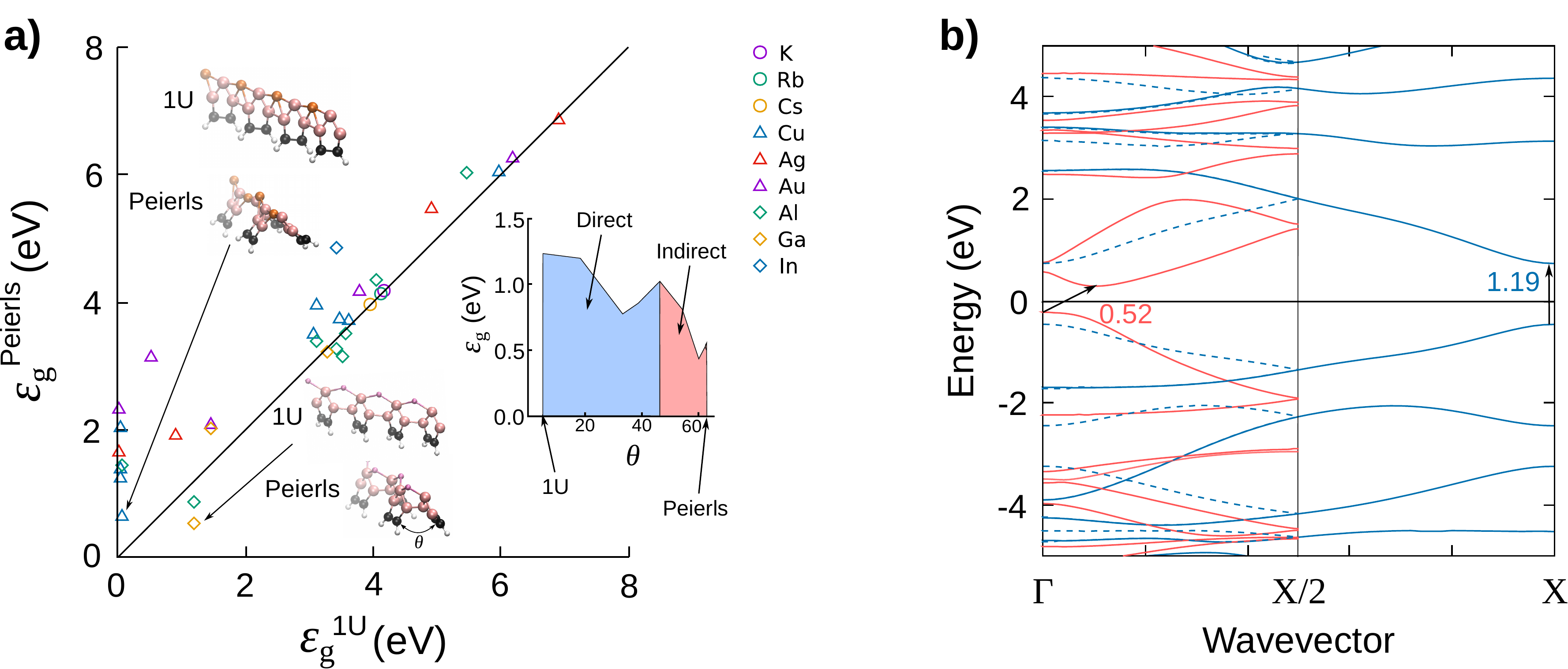}
    \caption{
Influence of Peierls stabilization on  
electronic band gap, $\varepsilon_{\rm g}$ for 33 selected Peierls materials:
a) Comparison of $\varepsilon_{\rm g}$, of 1U and Peierls phases.
Structures of gap-opening (with Cu) and gap-closing (with Ga) cases
containing the B$_3$C$_2$H$_2$ ring are shown. Variation in  
$\varepsilon_{\rm g}$ of B$_3$C$_2$H$_2$Ga during 1U-to-Peierls
geometry relaxation is shown in the subplot. 
b) Band folding and gap-closing in B$_3$C$_2$H$_2$Ga.
Band structures of 
1U (blue solid lines), 2U superstructure (blue dashed lines) and Peierls (red solid lines) phases
are shown. Arrows point to band gaps and Fermi energy level is set to 0 eV.}
    \label{fig:fig3}
\end{figure*}
Along the $I_1/I_3=1-I_2/I_3$ line in the shape distribution plot, one finds purely planar structures formed by metal ions $\sigma$-bonded to the rings. 
On the $I_2/I_3=1$ vertical line, one finds eclipsed 
$\eta^5:\eta^5$ sandwich structures with Cp and N$_5$ rings. 
The top-most point corresponds to
the most elongated sandwich complex N$_5$Rb;
while CpNa, at the bottom, is the most compact. 
In the latter, the lattice constant is more than twice the ring-metal distance; indicating
weak crystal packing leading to high-volatility similar to the experimentally studied CpGa\cite{downs1993chemistry}. These trends are in line with the ionic radii of the corresponding metals.

Structures with $\eta^2$/$\eta^3$/$\eta^4$-bonding arise predominantly for
hetero-substituted rings with cations slipped away from the geometric center of the ring. 
Experimentally, CpNa has been identified as a 1U system with 
an $\eta^5:\eta^5$ pattern\cite{dinnebier1997solid}, 
which is confirmed by our calculations (see Figure~\ref{fig:fig2}).  
However, the same metal in combination with 
substituted Cp rings shows a diverse coordination 
pattern due to lowering of 
the ring symmetry. Out of 1199 RMQ1D materials,
thermodynamically most and least stable 
are B$_2$NS$_2$H$_2$Cs (formation energy, $E_{\rm f}=-0.877$ eV), and 
B$_3$CNHCu ($E_{\rm f}=1.290$ eV), respectively; neither are dynamically stable in the 1U phase. 
The pentazole ring, N$_5^-$, isolable in an acidic medium\cite{zhang2019stabilization} 
prefers equatorial $\sigma$-bonding with cations\cite{christe2017polynitrogen}. 
Our results indicate hard acid ions like Cs$^+$ or Rb$^+$ 
to form N$_5^-$ complexes through $\eta^5:\eta^5$ coordination pattern. 
However, the same ring when complexed with the soft acid ions
Cu$^+$, Ag$^+$ and Au$^+$ forms $\eta^1:\eta^1$ $\sigma$-bonding. 
Thermodynamic stability of these materials correlates poorly with their 
dynamic stability, made apparent by the fact that $E_{\rm f}$ of the
38 materials stabilizing in the 1U phase uniformly span a window of 
$-0.726$ to $0.992$ eV.

RMQ1D materials that are neither unstable nor stable in the 1U phase comprise potential cases
for Peierls (587), CDW (28), and long wave (41) 
transitions. Peierls materials are detected by a phonon
instability at $q^*={\rm X}$ (see Figure~\ref{fig:fig1}). Given that the experimental evidence for RMQ1D materials stabilized by a Peierls transition
are far and few between, it is of interest to characterize those
preferring a low symmetric phase in our modeling. Presently it is not
computationally feasible to perform refined investigations of all 587 Peierls candidates.
Hence, we select the most important subsets including 16 materials with
 $|\overline{\nu}_{\rm im}| \ge50$ cm$^{-1}$
expected to show a large gain in the potential energy 
via soft-mode distortions.
Besides, we choose 10 Peierls candidates 
containing the rings Cp, B$_3$C$_2$H$_2^-$, or N$_5^-$; the
latter two have been selected to understand the impact of $\sigma-$aromaticity
on the geometry.
We also include 7 gap-opening Peierls 
candidates that are metallic in the 1U phase.
Starting from the 2U superstructure of the 1U geometries and a subsequent 
distortion along the
soft-mode of the 1U phase, a
full geometry relaxation 
was performed for these 33 materials.

Irrespective of the bandgap modulation, all 33 Peierls materials undergo structural distortions deviating from the 1U structure, augmented by a net drop in the total energy. 
A comparison of $\varepsilon_{\rm g}$ in the Peierls
and dynamically unstable 1U phases
reveals some interesting trends (Figure~\ref{fig:fig3}a). 
The magnitude of $\varepsilon_{\rm g}$ in the 
Peierls phase varies inversely with that of the 1U phase. Materials
with a large gap in the 1U phase ($\varepsilon_{\rm g}>3$ eV) show a small gain in the bandgap going to the Peierls phase. A common structural characteristic in all the gap-opening Peierls candidates is a slippage of the metal ions---especially Al, Cu, Ag, and Au---away from the ring center resulting in poor shielding of the metallic bonds by the
rings in the 1U phase. Figure S1 in the SI presents the relaxed structures of all 33 Peierls cases. 
Overall, 7 of these exhibit strong
gap-opening and 3 prefer an uncommon
gap-closing Peierls transition, $\varepsilon_{\rm g}^{\rm Peierls}<\varepsilon_{\rm g}^{1{\rm U}}$. 
\INBLUE{It is well-known that at finite-temperature, 
the Peierls structural distortion is lifted\cite{he2014finite}. In RMQ1D materials,
such a transition may be observed for the 33 
Peierls cases. Thermalization of the electronic states\cite{deshpande2010electron}  may be achieved with a large Fermi temperature for a frozen geometry. 
However, a more satisfactory dynamic picture may be obtained only through
atomistic modeling revealing a thermalized, 
symmetric lattice structure as experimentally noted for perovskites\cite{kirschner2019photoinduced}.}

Of all selected Peierls materials, most show a gap-opening transition
(Figure~\ref{fig:fig3}a). As an example of a material undergoing a gap-closing
transition, the structure of 
B$_3$C$_2$H$_2$Ga is shown in the inset of Figure~\ref{fig:fig3}a.
Compared to the bands in the 1U phase, those in the Peierls phase
are shifted down (Figure~\ref{fig:fig3}b). 
However, the doubly degenerate levels of the folded 1U bands at $E=2$ eV lie far above $E_{\rm F}$ to remove this degeneracy. 
An indirect bandgap in the Peierls phase is due 
to the energy ordering of bands and their alignments. The
dip in the lowest conduction band is due to an avoided crossing between the bands immediately below and above $E_{\rm F}$ showing mixed characteristics near the $\Gamma$-point. We verified this fact through an orbital analysis.
Dependence of the band mixing on distortion along the angle $\theta$ 
is illustrated in the inset to Figure~\ref{fig:fig3}a. 
For the 1U phase, $\theta$ is 4.43$^\circ$. 
\INBLUE{Since the character of the 
unstable phonon mode in the 1U phase is essentially 
angular, Peierls distortion
brings negligible change in the metal-metal distance, going from
3.76 \AA{} to 3.69 \AA{}.} 
During geometry relaxation of the 1U phase's 2U structure, $\theta$ converges to  $62.64^\circ$
at the Peierls phase. 
The characteristic of the bandgap
changes from direct to indirect at $\theta \thickapprox 45^\circ$.  
\begin{figure}
    \centering
    \includegraphics[width=7cm]{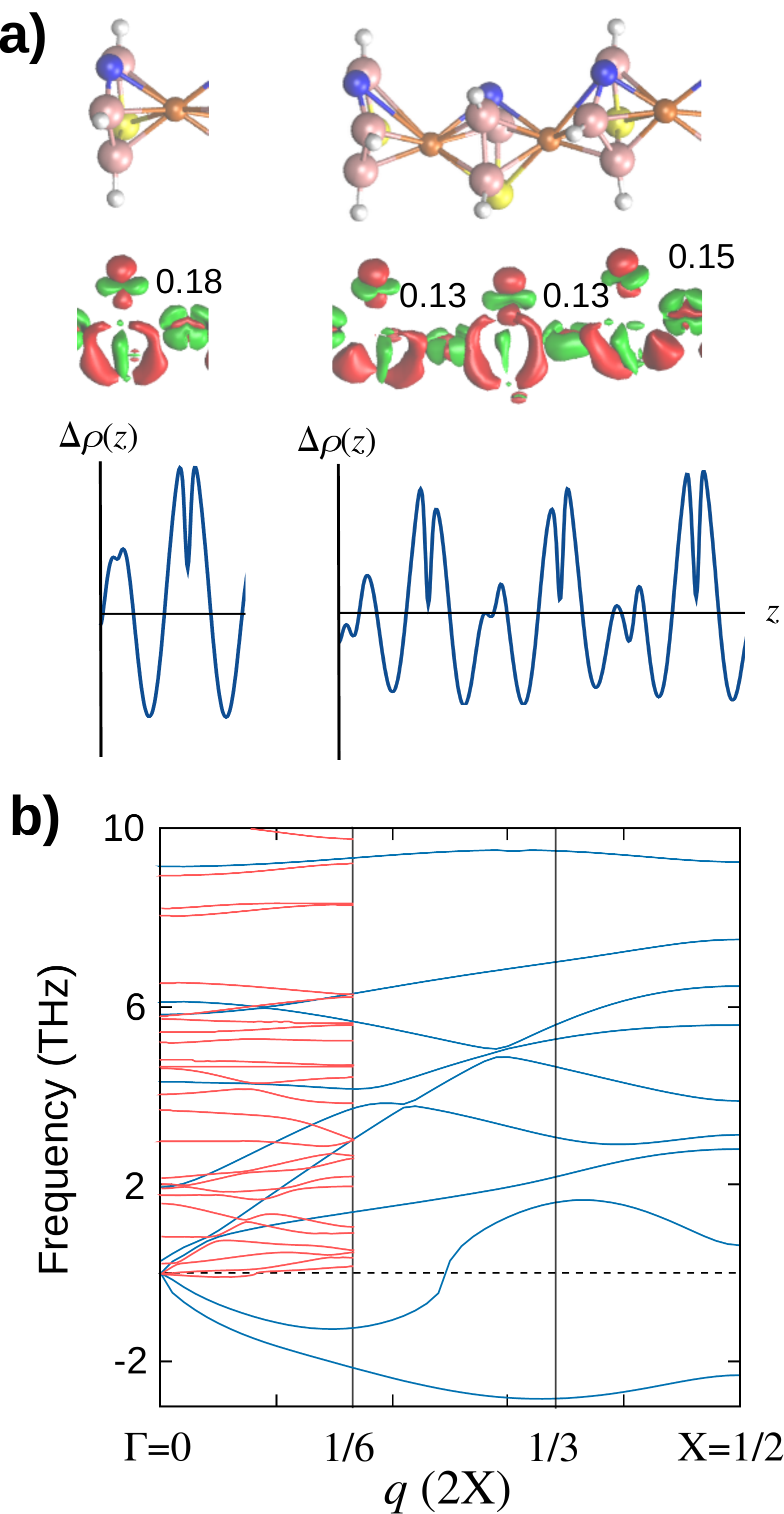}
    \caption{Charge density wave formation in B$_3$NSH$_3$Cu: 
    a) Unit cell arrangements in 1U (left) and CDW (right) phases are shown along with the density difference, $\Delta \rho({\bf r})$, calculated by subtracting the densities of rings and metals
    from the solid. 
    Green and red denote gain and depletion of electron density, respectively. 
    Mulliken charges on the Cu centers are stated in units of $e$.
    Also shown is the longitudinal density difference, $\Delta \rho(z)$ (blue solid line). 
    b) Phonon band structures of the 1U (blue lines) and the CDW (red lines) phases; the zone-edge
    X is $\pi/c$, where $c$ is the lattice constant. }
    \label{fig:fig4}
\end{figure}

Following an analysis of the 1U phonon spectrum of the 23 CDW candidates,
we select cases for which the magnitude of the maximally unstable
phonon mode is $\ge 50$ cm$^{-1}$ and the corresponding wavevector
coinciding with the commensurate values: 
$q^*=1/3$, $q^*=1/4$ and $q^*=1/5$, shown as a stick spectrum in Figure~\ref{fig:fig1}. Further analysis revealed
only 4 out of 23 to show commensurate phonon instability.
    Electron density of the 1U material is a periodic repetition of
its unit cell density, $\rho^{\rm u}$: 
    \begin{eqnarray}
\rho^{\rm 1U}(z) =  \rho^{\rm u} \cos( 2 {\rm X} z ).
\end{eqnarray}
In the commensurately modulated phase, the 1U density is distorted by a
periodic superstructure wave of amplitude $\Delta \rho^{\rm s}$ and
wavelength $1/q^*$
\begin{eqnarray}
\rho^{N\rm U}(z)=\rho^{\rm 1U}(z)+ \Delta \rho^{\rm s} \cos( z q^*).
\label{eq:cdw}
\end{eqnarray}
The only potential CDW material
satisfying our selection criteria is B$_3$NSH$_3$Cu
whose unit cell details and phonon band structures
are on display in Figure~\ref{fig:fig4}.
In the 1U phase, the metal centers are separated by 3.51 \AA{} with 
the ring shielding the interaction between adjacent metal centers rendering
the material an insulator,
$\varepsilon_{\rm g}^{1 {\rm U}}=1.16$ eV. While the 1U phonon band structure shows prominent instability at $q=2{\rm X}/3$, 
one also notes another band exhibiting instability 
at ${\rm X}/3$, albeit with phonon
wavenumber of a smaller magnitude.
In the 1U phase, a partial electron transfer from the anionic ring to the metal center is 
indicated by the net Mulliken charge on Cu, $q^{\rm M}_{\rm Cu}=0.18\,e$.
The density difference  
$\Delta \rho({\bf r})=\rho({\bf r})-\rho^{\rm rings}({\bf r})-\rho^{\rm metals}({\bf r})$ illustrates regions of electron density
gain and depletion upon ring-metal bonding in the solid. 
Better clarity is attained with the reduced 
 density difference, by integrating $\Delta \rho({\bf r})$ in the plane perpendicular to the lattice vector resulting in the
1D fingerprint, $\Delta \rho(z)=-\int \int dx  dy\,\Delta \rho({\bf r})$.
The shape and periodicity of $\Delta \rho(z)$ coincide with the 
structural and charge density variations of the material.

Full geometry relaxation based on the soft-mode distortion of the 3U superstructure was 
done as previously stated for the Peierls cases.
Transitioning to the dynamically stable CDW phase, the material trimerizes with a net drop
in the $E_{\rm f}$ by 0.031 eV.
Overall, the material shows further widening of the electronic
gap to 1.97 eV, the corresponding phonon spectrum confirms a dynamically stable phase (Figure~\ref{fig:fig4}b).  
The visible effect in the structure is a repeated chain of two 
distinct units: a metal-ring-metal trimer, with 2 equivalent Cu atoms ($q^{\rm M}_{\rm Cu}=0.13\,e$) which is 
displaced from a ring-metal-ring fragment along a transverse direction. 
Based on this arrangement, one can interpret the structure of B$_3$NSH$_3$Cu
through a `pseudo'-Peierls transition of the 3U superstructure phase. 
This structural description 
agrees with the periodic variation of 
$\Delta \rho({\bf r})$ and $q^{\rm M}_{\rm Cu}$, 
while the reduced density fingerprint
conveys further information that are not perceptible from the other quantities. 
Within the metal-ring-metal fragment, one notes  $\Delta \rho(z)$ coinciding with 
Cu to split in an asymmetric fashion. This is an indication of a weak bonding
 between the Cu atoms of metal-ring-metal, leaving the single Cu-center ($q^{\rm M}_{\rm Cu}=0.15\,e$) of ring-metal-ring 
in an environment similar to that of the 1U phase. 
While the amplitude of $\Delta \rho(z)$ is uniform 
across the lattice for the 1U phase,
it is undulatory for the CDW phase.

To conclude, we identify stable phases of RMQ1D materials based on maximum instability in the phonon spectrum of the translationally symmetric phase with one unit. 
Among those with multiple unstable phonon modes, there are potential candidates that
may be stabilized by a weak lateral packing in the bulk phase.
While our high-throughput workflow also identifies long wave candidates 
potentially distorting to a periodic structure containing $>5$ Us, 
it is 
beyond the scope of the present study to embark on their characterization. 
However, it may be
anticipated that the polymer design strategy based on a soft-phonon will be less reliable 
for the long wave cases because of the 
residual uncertainties associated with 
the methods used in resolving $q^*<0.2$. 
The procedures applied here can be extended to materials comprising
open-shell metal ions to identify candidates 
stabilized by magnetic moments, or half-metals relevant for spin-conduction\cite{wang2008novel}.
However, a  high-throughput study of magnetic RMQ1D will incur severe
computational complexities because of the challenges associated 
with the determination of the 
ground state among competing spin-states. 
For heavier atoms, besides, spin-orbit coupling needs to be incorporated for realistic modeling. \INBLUE{Extending the ideas presented here to higher-dimensional phases
requires the phonon wavevector to have more components. For instance, distortion
of the square-net phase of H atoms (1U phase) in 2D to 
a tetrameric phase (4 Us in the
unit cell)\cite{hoffman1988solids} is indicated by 
maximal phonon instability, 
${\bf q}^*=(0.5,0.5)$ in the 1U phase.}
The catalog of materials presented here can be further extended by considering other rings, metals, and their hetero-combinations. The results presented here and further explorations of them
through the publicly accessible interactive data-mining framework, may leverage
rational experimental design of Q1D materials
with tunable electronic properties.

\section*{Computational Details}
 Geometry optimizations of the anion rings were performed 
 using the program package Gaussian16\cite{g16short} with the 
PBE0\cite{adamo1999toward} density functional theory method 
and a def2-TZVP basis set. After eliminating 
unconverged, non-planar and acyclic structures,
the resulting 109 rings were combined with monovalent cations to construct
unit cell geometries for periodic calculations.
An initial lattice constant of minimum $c=5.0$ \AA{}---with the
metal placed at $c/2$ \AA{} from the ring---and
{\bf a} and {\bf b} vectors of length $50$ \AA~ were used to emulate vacuum.
Full relaxation of RMQ1Ds were performed with 
the all-electron, numeric atom-centered orbital code
FHI-aims\cite{blum2009ab} with the PBE\cite{perdew1996generalized} functional. In all calculations, we used
1x1x64 $k$-grids, and 
tight, tier-1 basis set for all atoms. 
Lattice vectors and atomic coordinates were fully relaxed 
with electron density converged to $10^{-6}$ $e/$\AA$^3$,
\INBLUE{analytic forces to $5\times10^{-4}$ eV/\AA},
and the maximum force 
component \INBLUE{for lattice relaxation} to $5\times10^{-3}$ eV/\AA. 
Phonon spectra were obtained \INBLUE{with a 113 supercell} using 
finite-derivatives of analytic forces
with an atomic displacement of 0.005 \AA{} 
and tighter thresholds \INBLUE{($10^{-7}$ $e/$\AA$^3$ for density;
$10^{-6}$ eV/\AA{} for analytic forces)}
using Phonopy\cite{togo2015first}
interfaced with FHI-aims. \INBLUE{These control settings yield
converged results as shown in Table~S2 and Table~S3 in the SI.
Table~S1 compares the performance of the PBE functional
with other semilocal and hybrid ones. Since the packing interaction in the
RMQ1D materials is essentially ionic in nature, van der Waals 
correction to PBE influences the
geometries and phonons negligibly. The lack of critical long-range
interactions in the RMQ1D materials is also indicated by the fact that
the PBE results are comparable to that of the hybrid 
methods HSE03 or PBE0. 
Treatments at the latter level are necessary for 
accurate modeling of Peierls phases with significant
long-range interaction within the unit cell as noted for polyacetylenes\cite{dumont2010peierls}.}
Single point calculations were
performed with the PBE0 functional
for accurate estimations of total energies and band gaps. 
 
\section*{Supplementary Material}
The supplementary material contains: 
(i) additional tables 
with results benchmarking the performance of
 DFT methods and  control parameters
for selected RMQ1D materials,
(ii) reference chemical potential for calculating formation energies, and
(iii) screenshots for data-mining (see Data Availability).

\section*{Acknowledgments}
RR gratefully acknowledges Prof. Matthias Scheffler for providing a license to the FHI-AIMS program. 
The authors thank Salini Senthil for setting up the data-mining framework.
We acknowledge support of the Department of Atomic Energy, Government
of India, under Project Identification No.~RTI~4007.
All calculations have been performed using the Helios computer cluster, which is an integral part of the MolDis Big Data facility, TIFR Hyderabad \href{https://moldis.tifrh.res.in}{(https://moldis.tifrh.res.in)}.

\section*{Data Availability}
The data that support the findings of this study are openly available
at \href{http://moldis.tifrh.res.in/data/rmq1d}{(https://moldis.tifrh.res.in/data/rmq1d)}. 
\INBLUE{
Input and output files of corresponding calculations are deposited in the NOMAD repository \href{https://nomad-lab.eu/}{(https://nomad-lab.eu/)}.}
\bibliography{lit} 
\end{document}